\documentclass[]{spie}  

\usepackage{amsmath,amsfonts,amssymb}
\usepackage{float}
\usepackage{graphicx}
\usepackage[colorlinks=true, allcolors=blue]{hyperref}

\title{An effective interactive brain cytoarchitectonic parcellation framework using pretrained foundation model}

\author[1,2,3]{Shiqi Zhang}
\author[2,3]{Fang Xu}
\author[2,3]{Pengcheng Zhou} 
\affil[1]{Department of Biomedical Engineering, Southern University of Science and Technology, Shenzhen, 518055, Guangdong, China}
\affil[2]{Interdisciplinary Center for Brain Information, Brain Cognition and Brain Disease Institute, Shenzhen Institutes of Advanced Technology, Chinese Academy of Sciences, Shenzhen, China}
\affil[3]{Faculty of Life and Health Sciences, Shenzhen University of Advanced Technology, Shenzhen, 518107, Guangdong, China}

\authorinfo{Further author information: (Send correspondence to Pengcheng Zhou)\\
Shiqi Zhang.: E-mail: sq.zhang1@siat.ac.cn \\
Fang Xu.: E-mail: xufang@suat-sz.edu.cn \\
Pengcheng Zhou.: E-mail: zhoupc@suat-sz.edu.cn}

\begin{document}
\maketitle

\begin{abstract}
Cytoarchitectonic mapping provides anatomically grounded parcellations of brain structure and forms a foundation for integrative, multi-modal neuroscience analyses. These parcellations are defined based on the shape, density, and spatial arrangement of neuronal cell bodies observed in histological imaging. Recent works have demonstrated the potential of using deep learning models toward fully automatic segmentation of cytoarchitectonic areas in large-scale datasets, but performance is mainly constrained by the scarcity of training labels and the variability of staining and imaging conditions. To address these challenges, we propose an interactive cytoarchitectonic parcellation framework that leverages the strong transferability of the DINOv3 vision transformer. Our framework combines (i) multi-layer DINOv3 feature fusion, (ii) a lightweight segmentation decoder, and (iii) real-time user-guided training from sparse scribbles. This design enables rapid human-in-the-loop refinement while maintaining high segmentation accuracy. Compared with training an nnU-Net from scratch, transfer learning with DINOv3 yields markedly improved performance. We also show that features extracted by DINOv3 exhibit clear anatomical correspondence and demonstrate the method’s practical utility for brain region segmentation using sparse labels. These results highlight the potential of foundation-model-driven interactive segmentation for scalable and efficient cytoarchitectonic mapping. Code can be found on \href{https://github.com/Confetti22/cytoarch_brain_parcellation_dino}{https://github.com/Confetti22/cytoarch\_brain\_parcellation\_dino}.

\end{abstract}


\begin{figure}[htbp]
\begin{center}
\begin{tabular}{c}
\includegraphics[width=0.8\linewidth]{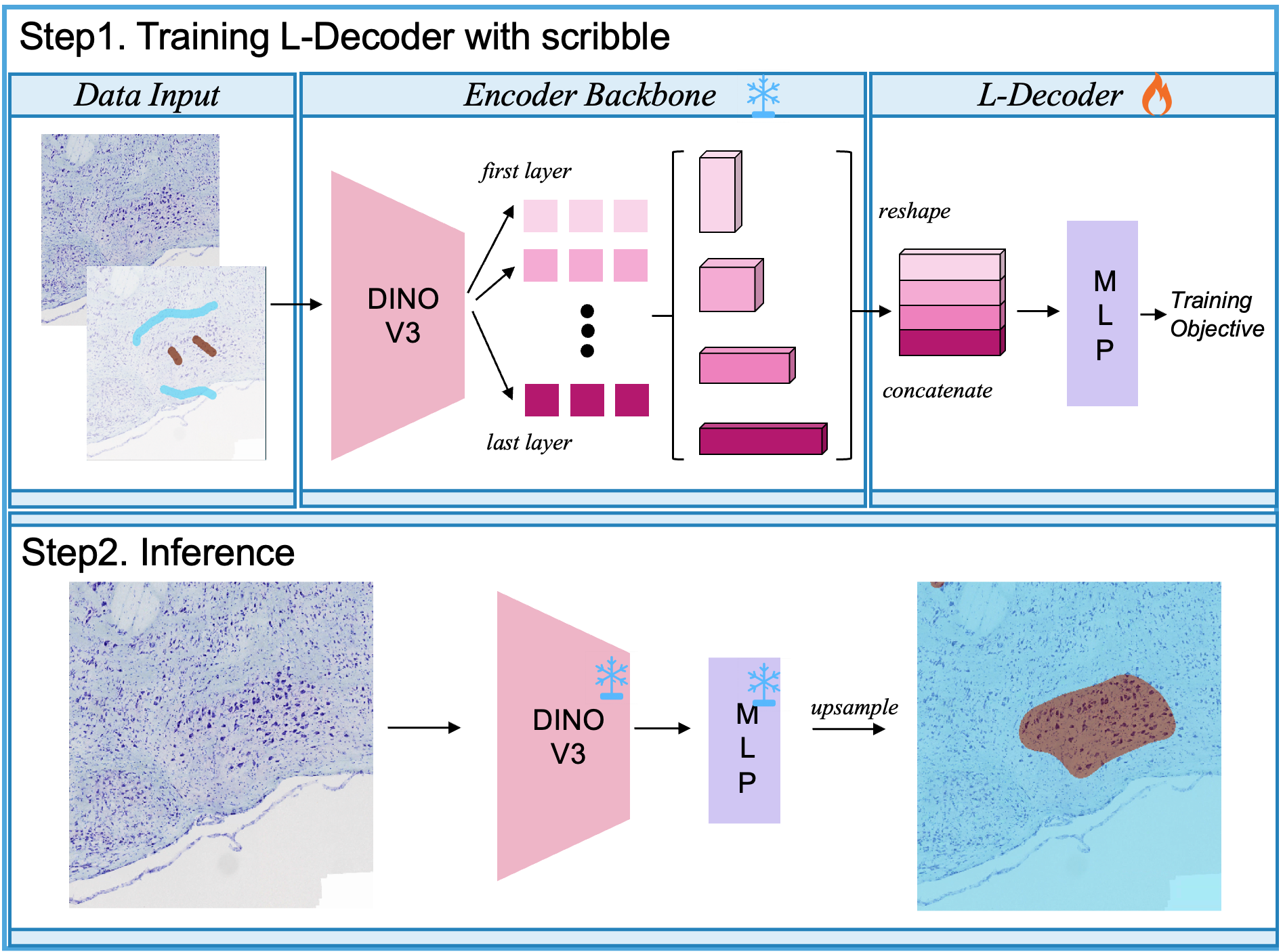} 
\end{tabular}
\end{center}
\caption{Overview of the proposed interactive cytoarchitectonic parcellation framework. A pretrained DINOv3-B vision transformer serves as a frozen encoder. Multi-layer features from different depths are upsampled, aligned, and concatenated, then passed to a lightweight MLP head to produce the final segmentation mask. Sparse user-provided scribbles are used to fine-tune the decoder, enabling rapid, interactive parcellation.}
\label{fig:framework}
\end{figure}

\section{INTRODUCTION}
\label{sec:intro}

Cytoarchitectonic parcellation provides a critical structural basis for understanding brain organization by delineating regions according to the spatial arrangement and morphology of neuronal cell bodies in histological sections \cite{amunts2013bigbrain}. Such parcellations support a broad range of neuroscientific applications, including functional annotation, atlas construction, and multimodal data integration. However, creating accurate cytoarchitectonic maps typically requires extensive manual annotation by trained neuroanatomists, which is slow, labor-intensive, and difficult to scale to large datasets.

Deep learning is a powerful paradigm because it can learn data representations that substantially simplify downstream tasks. When high-quality features are available, even simple classifiers or segmentation heads can achieve strong performance. However, in practice, training effective feature extractors is severely constrained by the scarcity of expert annotations and the substantial variability across histological modalities. Without sufficient labeled data, conventional models struggle to learn representations that generalize across stains, species, and imaging protocols.

In contrast, large vision foundation models trained on massive natural-image corpora offer a promising alternative. These models already possess strong, transferable feature representations learned in a self-supervised manner, which can often be adapted to new domains with minimal supervision. Building on this insight, we treat the foundation model as a fixed feature extractor and introduce a lightweight decoder that can be trained directly from sparse user-provided scribbles. This combination enables an efficient human-in-the-loop workflow, allowing interactive cytoarchitectonic segmentation with only minimal annotation effort.

In this work, we introduce an interactive cytoarchitectonic parcellation framework that leverages DINOv3 as a frozen feature extractor and uses a lightweight decoder trained directly from sparse scribble annotations. This framework integrates multi-layer feature fusion, fast decoder optimization, a tiled-inference strategy that processes large histological images, and a Napari-based interactive interface into a unified human-in-the-loop workflow. This design allows expert users to guide and refine parcellations within seconds, enabling practical and scalable annotation across diverse datasets.

We make the following contributions:
\begin{itemize}
    \item We demonstrate substantial improvements over nnU-Net on macaque V1 laminar segmentation.
    \item We show that DINOv3 features exhibit clear anatomical correspondence across stains and species.
    \item We implement a scribble-based interactive segmentation tool for cytoachitecture brain parcellation.
\end{itemize}
Together, these results highlight the potential of a foundation model-based transfer learning approach to transform cytoarchitectonic parcellation workflows.

\section{RELATED WORK}
\label{sec:related}

\subsection{Automatic methods for brain region parcellation}

Recently, deep-learning-based methods have been utilized to achieve automatic brain mapping for large-scale datasets. Spitzer et al.\cite{spitzer2017parcellation} modified the U-Net architecture to handle large image patches of size $2000\times2000$ pixels for classification of cytoarchitectonic areas in the human brain, but performance suffers from the limited amount of expert annotations. To circumvent this limitation, Schiffer et al.\cite{schf2021contra} use supervised contrastive learning as a pretraining method to improve downstream cytoarchitectonic classification performance. Although the pretrained model outperforms training from scratch, the overall classification precision is still not satisfactory. In this work, we achieve a more efficient and general-purpose parcellation method by utilizing the transferability of vision foundation models, which only requires sparse annotation.

\subsection{Leveraging the transferability of vision foundation models}

Foundation models demonstrate strong cross-domain transferable capabilities. Through large-scale pretraining, they acquire representations with inherent inductive biases that enhance out-of-distribution generalization, especially under limited-data fine-tuning\cite{fort2021exploring,hendrycks2020pretrained}. Recent works have exploited this advantage for microscopy bioimaging, where annotation is expensive to acquire. Chen et al.\cite{chen2019active} use Inception-BN pretrained on ImageNet to extract texture features for brain-stem nuclei classification in mouse brain and achieve human-comparable precision. However, Inception-BN's classical CNN features capture broad histological texture but lack accurate local features, which hinders its application to finer structures and more challenging situations where changes are subtle, such as laminar structure segmentation.

Prolonged and large-scale training can usually lead to improvements on global benchmarks but a degradation on dense tasks; this phenomenon is caused by the loss of patch-level consistency over training. DINOv3 is the first to tackle this issue via Gram anchoring, which results in superior features that are excellent at both high-level semantic tasks and precise dense prediction \cite{simeoni2025dinov3}. We utilize this advantage and use DINOv3 as a backbone to extract rich features from cellular-resolution brain imaging. Ranftl et al. \cite{ranftl2021vision} design a lightweight decoder with multi-level feature fusion to effectively adapt the representations from the DINO family for segmentation. We take up this idea with sparse user input, implementing interactive segmentation.

\setlength{\tabcolsep}{3pt}
\begin{table}[ht]
\centering
\small
\caption{Comparison with nnU-Net for cortical layer segmentation.}
\label{metrics}
\begin{tabular}{|l|c|c|c|c|c|c|c|c|c|c|c|c|}
\hline
Class & \multicolumn{2}{c|}{DSC} & \multicolumn{2}{c|}{IoU} & \multicolumn{2}{c|}{Rec.} & \multicolumn{2}{c|}{Prec.} & \multicolumn{2}{c|}{HD95 (um)} & \multicolumn{2}{c|}{ASSD (um)} \\
 & Ours & nnU-Net & Ours & nnU-Net & Ours & nnU-Net & Ours & nnU-Net & Ours & nnU-Net & Ours & nnU-Net \\
\hline
L1 & \textbf{0.796} & 0.544 & \textbf{0.721} & 0.480 & \textbf{0.924} & 0.846 & \textbf{0.777} & 0.545 & \textbf{87.947} & 324.458 & \textbf{30.005} & 89.706 \\
\hline
L2/3 & \textbf{0.733} & 0.530 & \textbf{0.654} & 0.437 & \textbf{0.902} & 0.591 & \textbf{0.722} & 0.651 & \textbf{251.042} & 507.981 & \textbf{81.601} & 182.628 \\
\hline
L4A & \textbf{0.606} & 0.424 & \textbf{0.479} & 0.302 & \textbf{0.715} & 0.570 & \textbf{0.622} & 0.393 & \textbf{251.382} & 696.554 & \textbf{88.803} & 253.866 \\
\hline
L4B & \textbf{0.666} & 0.453 & \textbf{0.547} & 0.326 & \textbf{0.756} & 0.549 & \textbf{0.675} & 0.448 & \textbf{231.844} & 680.654 & \textbf{83.667} & 224.358 \\
\hline
L4Ca & \textbf{0.604} & 0.367 & \textbf{0.492} & 0.254 & \textbf{0.694} & 0.391 & \textbf{0.635} & 0.433 & \textbf{260.679} & 615.900 & \textbf{90.196} & 208.470 \\
\hline
L4Cb & \textbf{0.552} & 0.347 & \textbf{0.420} & 0.237 & \textbf{0.557} & 0.497 & \textbf{0.620} & 0.339 & \textbf{201.334} & 636.462 & \textbf{71.107} & 188.748 \\
\hline
L5 & \textbf{0.676} & 0.389 & \textbf{0.580} & 0.288 & \textbf{0.777} & 0.537 & \textbf{0.706} & 0.416 & \textbf{258.934} & 756.876 & \textbf{79.790} & 249.217 \\
\hline
L6 & \textbf{0.482} & 0.317 & \textbf{0.399} & 0.232 & \textbf{0.720} & 0.404 & \textbf{0.509} & 0.401 & \textbf{448.231} & 1013.283 & \textbf{135.064} & 342.269 \\
\hline
overall & \textbf{0.639} & 0.425 & \textbf{0.536} & 0.323 & \textbf{0.756} & 0.546 & \textbf{0.658} & 0.456 & \textbf{248.924} & 645.109 & \textbf{82.529} & 215.288 \\
\hline
\end{tabular}
\end{table}

\begin{figure}[htbp]
\begin{center}
\begin{tabular}{c}
\includegraphics[width=0.8\linewidth]{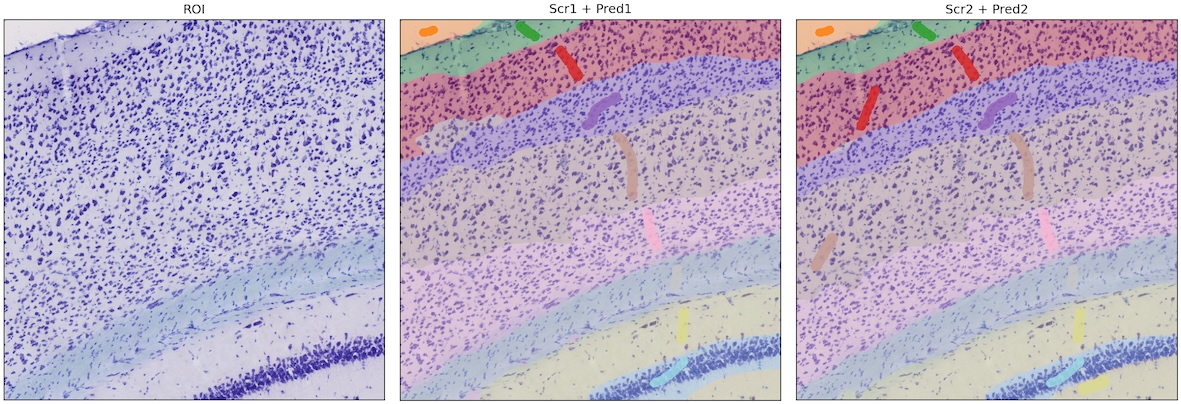} 
\end{tabular}
\end{center}
\caption{Interactive segmentation, in which segmentation accuracy is progressively improved through iterative refinement of user annotations.}
\label{fig:inter_step}
\end{figure}

\section{METHODS}
\label{sec:methods}

Our goal is to develop a flexible, interactive cytoarchitectonic parcellation method capable of learning from sparse scribble annotations. The major challenge lies in the scarcity of labeled training data and substantial variations in imaging conditions across datasets. To address this, we exploit the exceptional transferability of DINOv3 to enable few-shot training with minimal user input.

\subsection{Multi-layer DINOv3 feature fusion with a lightweight decoder}
\label{sec:feature-fusion}

To effectively use knowledge learned by DINOv3, we fuse multi-layer features from the DINOv3 backbone and pass them to a lightweight decoder to produce the final segmentation mask, inspired by the light-decoder design in Ranftl et al.\cite{ranftl2021vision}.

We use a pretrained DINOv3 Vision Transformer (ViT) as the encoder and keep all its weights frozen during training. Given an input image, the encoder splits it into non-overlapping patches, each mapped to a $d$-dimensional token, then processes these tokens through $L$ Transformer blocks, producing a sequence of hidden states with progressively richer semantic content.

To capture both low-level and high-level semantics, we extract patch tokens from a small set of intermediate layers (e.g., lower, middle, and higher blocks). For each layer, only patch tokens are retained; auxiliary tokens such as the class or register tokens are discarded. The selected tokens are then reshaped into feature maps and projected through a $1\times1$ convolution to unify the channel dimension and then refined with a $3\times3$ convolution to enhance local spatial coherence. All maps are then upsampled to a unified spatial resolution and concatenated along the channel dimension to form a fused representation.

A compact MLP decoder then maps the fused multi-level feature to per-patch class predictions, producing the final segmentation mask. This lightweight design keeps training efficient while maintaining strong capacity for dense prediction.

\subsection{Loss design under class imbalance}
\label{sec:loss}

Cytoarchitectonic regions are often highly imbalanced, especially thin cortical layers. To address this, we use a combined focal cross-entropy \cite{lin2017focal}and Dice loss with an additional $L_2$ weight regularization term. The focal component down-weights easy samples and emphasizes hard or minority classes, while the Dice component optimizes region overlap and improves boundary fidelity for thin structures, together with an $L_2$ weight regularization that regularizes all weights and biases as follows:

\begin{equation}
\mathcal{L}_{\text{total}} = \mathcal{L}_{\text{focal CE}} + \lambda_{\text{dice}} \mathcal{L}_{\text{Dice}} + \lambda_{w} \sum_{i} \lVert \theta_i \rVert_2^2,
\end{equation}

\noindent
where $\theta_i$ denotes the trainable parameters of the decoder, $\mathcal{L}_{\text{focal CE}}$ is the focal cross-entropy loss, and $\mathcal{L}_{\text{Dice}}$ is the soft Dice loss. (Exact formulas can be inserted here.)

\subsection{Overlapped patch inference}
\label{sec:overlap-inference}

To enable inference on large histological images that exceed GPU memory limits, we use an overlapped patch inference strategy combined with a radial blending mask. The large region of interest (ROI) is tiled into overlapping patches, each independently processed by the DINOv3 encoder and lightweight decoder. Before recomposition, each patch prediction is multiplied by a radially symmetric blending mask, which assigns maximal weight to the tile center and smoothly decreases toward the edges as a function of radial distance. Overlapping predictions are accumulated and normalized by the sum of their blending weights, producing a seamless, artifact-free output.

\begin{figure}[ht]
\begin{center}
\begin{tabular}{c}
\includegraphics[width=0.8\linewidth]{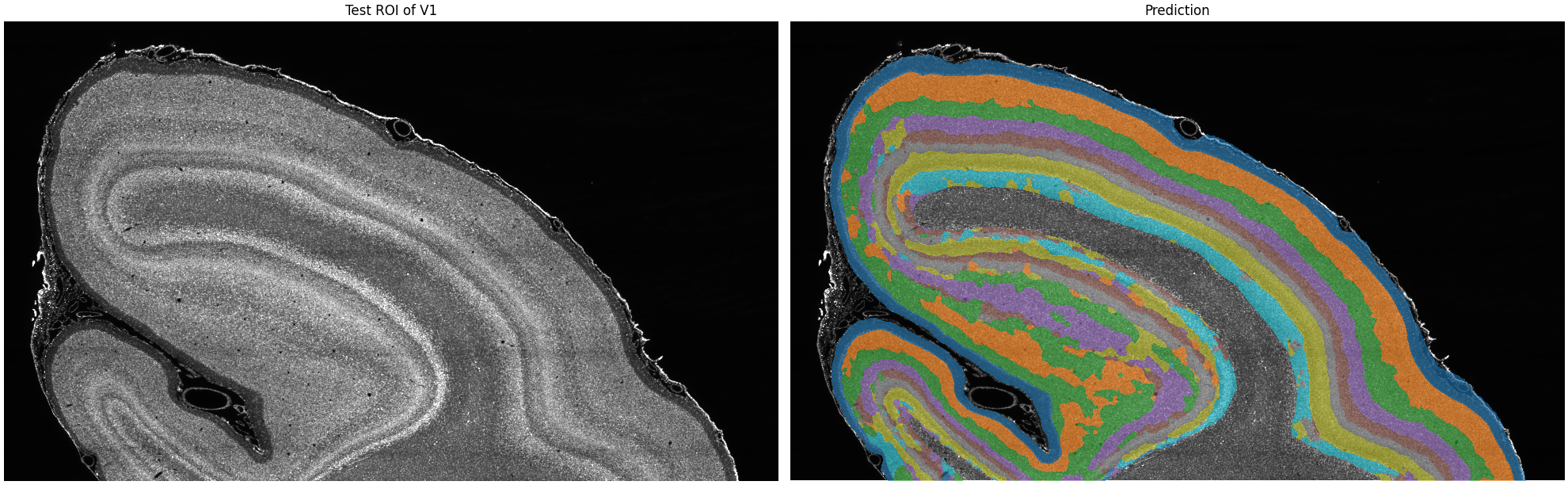} 
\end{tabular}
\end{center}
\caption{Predicted V1 cortical laminar segmentation on a test section. The method produces accurate and smooth layer boundaries in vertical regions where laminar structure is clearly preserved. }
\label{fig:v1}
\end{figure}

\subsection{Interactive segmentation via napari GUI}
To enable real-time user-guided parcellation, we integrate our model into a Napari-based interactive GUI. Owing to the high-quality texture representations encoded by DINOv3 and the small number of trainable parameters in the lightweight decoder, the model can be fine-tuned from sparse scribbles within only a few seconds, this allows users to iteratively annotate the region where model prediction is not satisfactory and immediately update the segmentation result, forming an efficient human-in-the-loop workflow(Fig.~\ref{fig:inter_step}), where users progressively refine labels and obtain immediate feedback. Specifically, the user first draws color-coded scribbles for different target regions on a predefined label layer. The training and inference pipeline is then triggered via a shortcut key, and the updated segmentation is rendered in the GUI within seconds. Guided by the displayed result, the user can revise existing annotations or add additional scribbles in regions with inaccurate or ambiguous predictions, and subsequently retrain the model using the refined scribbles to obtain more accurate parcellation outcomes, repeating this process until a satisfactory result is achieved.

\section{IMPLEMENTATION DETAILS}
\label{sec:implementation}

\subsection{V1 laminar segmentation dataset}
\label{sec:v1-dataset}

We use Nissl-stained rhesus macaque brain sections imaged with VISoR\cite{xu2021high} at $1\times1\times2.5~\mu\mathrm{m}$ resolution. Only high-quality non-oblique slices where the laminar structure is clearly visible are retained and downsampled to $4~\mu\mathrm{m}$ for computational efficiency. We label the 8-layer structure from V1, which includes Layer I (L1), Layer II/III (L2/3), Layer IVA (L4A), Layer IVB (L4B), Layer IVC$\alpha$ (L4Ca), Layer IVC$\beta$ (L4Cb), Layer V (L5), and Layer VI (L6). For training, 1614 patches ($512\times512$ pixels at $4~\mu\mathrm{m}$ per pixel) in V1 are extracted from the first 87\% of the sections, and 193 patches from the last 13\% are reserved for testing.

\subsection{Feature visualization and sparse-label datasets}
\label{sec:datasets}

We additionally consider three datasets for feature visualization and sparse-label interactive segmentation, and evaluate model performance at both $4~\mu\mathrm{m}$ (downsampled) and $1~\mu\mathrm{m}$ resolutions.

\begin{itemize}
    \item \textbf{Dataset 1:} A set of coronally cut sections of mouse brain stained with Nissl and imaged using VISoR at a $1\times1\times2.5~\mu\mathrm{m}$ resolution. We use reconstructed coronal slices at $1~\mu\mathrm{m} \times 1~\mu\mathrm{m}$ in-plane resolution.
    \item \textbf{Dataset 2:} A set of coronally cut sections of mouse brain stained with DAPI and imaged using a whole-slide scanner at $0.5~\mu\mathrm{m}$ resolution.
    \item \textbf{Dataset 3:} A set of sagittally cut sections of mouse brain at a thickness of $20~\mu\mathrm{m}$ that are stained with thionin and imaged in brightfield at $0.46~\mu\mathrm{m}$ resolution.
\end{itemize}

\begin{figure}[H]
\begin{center}
\includegraphics[width=1.0\linewidth, height=0.85\textheight, keepaspectratio]{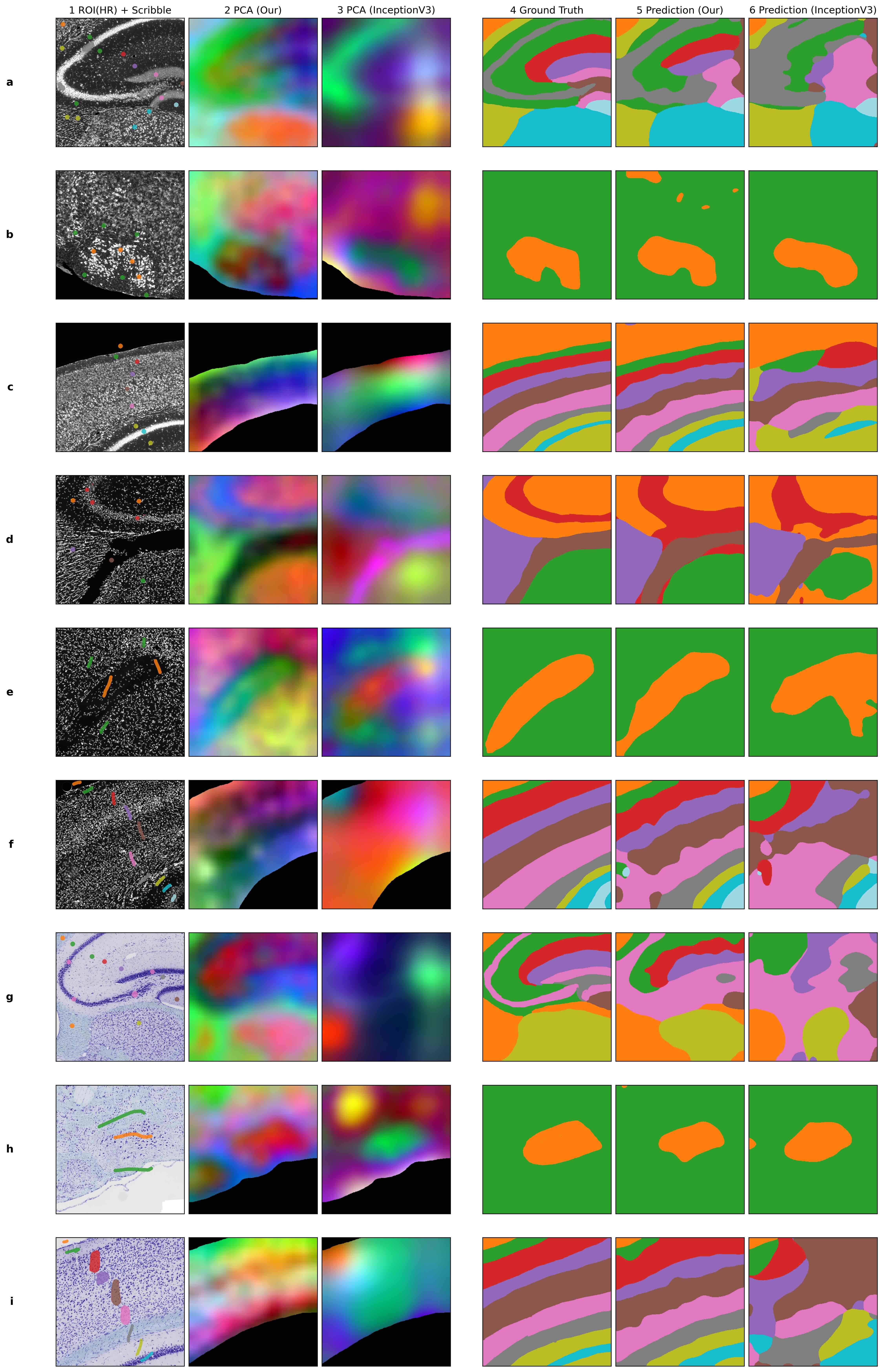}
\end{center}
\caption{Feature visualization and interactive segmentation results across three datasets. (a–c) Dataset 1 (Nissl-stained VISoR mouse brain), (d–f) Dataset 2 (DAPI-stained whole-slide images), and (g–i) Dataset 3 (thionin-stained sagittal sections). PCA projection of DINOv3 embeddings yields RGB maps with sharp anatomical boundaries. Scribble-based interactive segmentation results are shown for representative regions including Hippocampal formation (HIP), Facial Motor Nucleus (VII), Facial Nerve (VIIn), and Visual Cortex (VIS). For nucleus-stained data (Dataset 2), VIIn replaces VII due to reduced visibility of neuronal somata. Cytoarchitectonically similar structures are grouped when distinctions are subtle even for human experts (e.g., stratum oriens vs. stratum radiatum in HIP).}
\label{fig:vis}
\end{figure}

\subsection{Training settings}
\label{sec:training}

We optimize the model using the AdamW optimizer with a learning rate of $5\times10^{-4}$ and a weight decay of $1\times10^{-4}$. Training is performed for 50 epochs with a batch size of 16 (SGD-like schedule with constant learning rate for the decoder parameters). The DINOv3 backbone is kept frozen throughout training, and only the lightweight decoder parameters are updated. We employ the DINOv3-B encoder and extract intermediate representations from the 3rd, 6th, 9th, and 12th Transformer blocks to construct the multi-level feature set used by the decoder. The $\lambda$ parameter in the combo loss is set to 0.33 empirically.

To reduce high-frequency noise and staining artifacts common in histology, we apply a total variation (TV) regularizer to the decoder output. TV encourages spatial smoothness while preserving boundaries, resulting in cleaner and more stable segmentation masks. We use \texttt{denoise\_tv\_chambolle} provided by \texttt{scikit-image}.

\section{EXPERIMENTS}
\label{sec:experiments}

To investigate the benefit of transfer learning for cytoarchitectonic parcellation, we compare the performance of our method with nnU-Net on a V1 cortical layer segmentation task. nnU-Net is a widely used model architecture in the biomedical imaging field, and Wang et al.~\cite{wang2024deep} have successfully used it to segment laminar structure of the auditory cortex in the brain of the common marmoset monkey. For evaluation metrics, we report both overlap-based (DSC, IoU, precision, recall) and distance-based metrics (HD95, ASSD). This combination provides a comprehensive assessment of region-level overlap and boundary fidelity. Furthermore, we demonstrate that the knowledge learned from natural images by DINOv3 can extract meaningful features from histology brain images that can distinguish several cytoarchitectonic areas.

\subsection{V1 cortical layer segmentation}
Accurate V1 layer segmentation is essential for understanding canonical cortical microcircuit architecture and serves as a benchmark for evaluating fine-grained cytoarchitectonic segmentation. Following Wang et al.~\cite{wang2024deep}, we adjust nnU-Net’s configuration to accommodate $512\times512$ patches to ensure enough context information for lamina segmentation, and set the data-split policy to use our own train–test split. The nnU-Net is trained for 50 epochs using the standardized nnU-Net v2 pipeline. The model architecture, data preprocessing, and training strategies are all configured by the package itself.

In all evaluation metrics, our method achieves substantially higher scores than nnU-Net, as shown in Table~\ref{metrics}. These results demonstrate that DINOv3 transfer learning provides a significant advantage in scenarios with sparse and fine-grained annotations.

\subsection{Visualization of features extracted by DINOv3}
\label{sec:feature-vis}

To better understand the representations learned by DINOv3, we visualize its patch-level embeddings across multiple datasets. High-dimensional token embeddings are reduced to three principal components using PCA and mapped to the RGB space to form intuitive feature maps. These visualizations consistently exhibit sharp anatomical boundaries, low noise, and strong correspondence with the underlying cytoarchitectonic organization.

We compare these feature maps with those obtained from Inception-BN using the same PCA visualization procedure. DINOv3 produces substantially more coherent and discriminative representations (Fig.~\ref{fig:vis}), especially in regions where subtle texture cues define the borders between adjacent cytoarchitectonic areas.

\begin{figure}[ht]
\begin{center}
\begin{tabular}{c}
\includegraphics[width=0.8\linewidth]{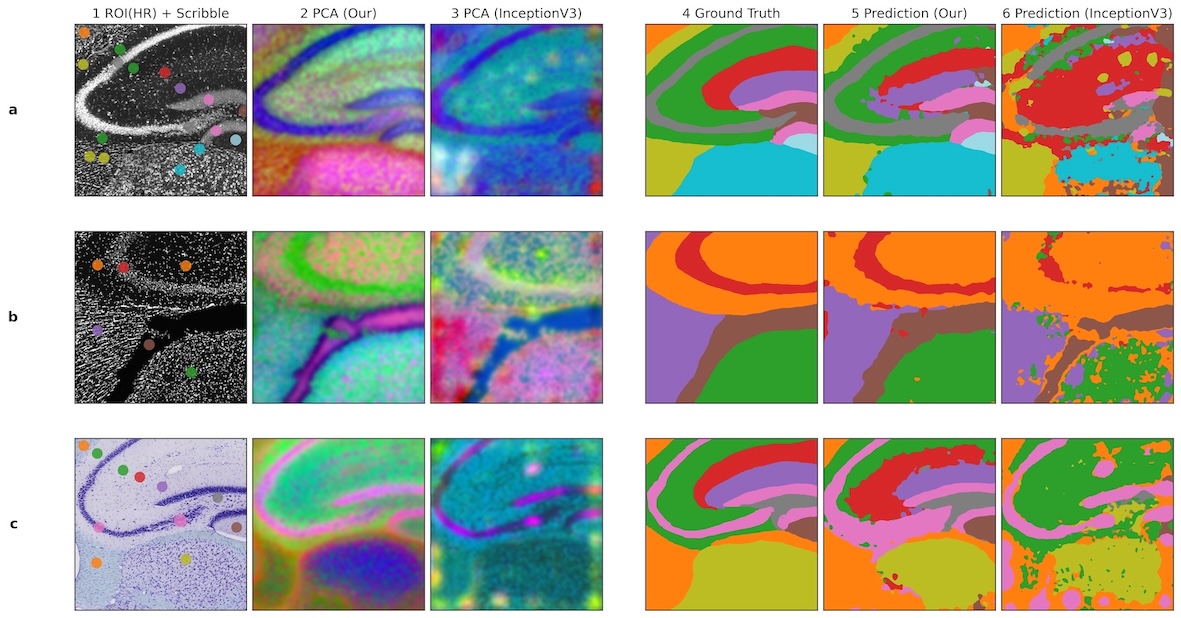} 
\end{tabular}
\end{center}
\caption{High-resolution ($1~\mu$m) DINOv3 feature visualizations and interactive segmentation. Hippocampal regions from the three datasets are processed at higher input resolution, revealing finer laminar and subfield distinctions. DINOv3 embeddings become more discriminative at $1~\mu\mathrm{m}$ resolution, enabling accurate segmentation of thin structures such as the pyramidal layer and improved separation of subtle subregions including stratum lacunosum-moleculare and the dentate molecular layer.}
\label{fig:hip-highres}
\end{figure}

To further assess the benefit of resolution, we perform the same analysis at higher input resolution ($1~\mu\mathrm{m}$). In hippocampal regions, DINOv3 embeddings reveal fine laminar distinctions—including thin pyramidal layers and subtle transitions between subfields—that are difficult to discern with conventional CNN features (Fig.~\ref{fig:hip-highres}). These results illustrate that DINOv3 not only captures global histological structure but also encodes fine-grained textural cues essential for precise cytoarchitectonic parcellation.

\subsection{Scribble-supervised segmentation}
\label{sec:napari}
To evaluate the effectiveness of this interactive setup, we compare our method with an Inception-BN–based baseline under identical conditions. Given a user-provided scribble mask, we first crop the corresponding $512\times512$ ROI which contains scribbles and use it as the training patch set. Due to sparse supervision, the decoder is fine-tuned for 15 epochs using the same training strategy employed for V1 laminar segmentation. For the Inception-BN baseline, we extract hierarchical convolutional features from the stem, Mixed\_5, Mixed\_6, and Mixed\_7 blocks, fuse them in the same manner, and attach the identical lightweight decoder for a fair comparison.

Across all scenes in the three datasets, our method consistently produces more accurate and less noisy segmentations than the Inception-BN baseline. Improvements are particularly pronounced for fine-grained structures such as cortical layers and the layered organization of the hippocampus. The advantage persists at higher input resolutions ($1~\mu\mathrm{m}$), where thin laminar structures (e.g., the pyramidal layer) are sharply delineated by our model. Subregions with subtle cytoarchitectonic differences—such as the stratum lacunosum-moleculare of Ammon’s horn and the molecular layer of the dentate gyrus—are also reliably distinguished. These results demonstrate that combining DINOv3 features with lightweight, scribble-based fine-tuning yields a powerful and practical tool for interactive cytoarchitectonic parcellation.

\section{DISCUSSION AND CONCLUSION}
\label{sec:discussion}

We presented a foundation-model-based method for interactive brain cytoarchitectonic parcellation that requires only sparse user supervision. By leveraging the transferable representations of DINOv3 and a lightweight decoder, the method performs rapid, accurate segmentation with minimal annotation effort. Compared with nnU-Net, our approach delivers superior overlap and boundary accuracy on the V1 laminar segmentation task. The ability of DINOv3 to capture fine-grained anatomical structure is evident in both feature visualizations and segmentation results across diverse datasets.

The interactive nature of our system makes it particularly suited for large-scale neuroanatomical mapping, where manual annotation capacity is limited and fully automatic method are sensitive to data shift. Our method generalizes well across different imaging modalities and resolutions, enabling flexible and efficient cytoarchitectonic parcellation under sparse supervision.

Future directions include applying this framework to whole-brain datasets, integrating multi-modal stains, and exploring cross-species feature alignment. Overall, our results demonstrate that leveraging DINOv3’s transferable representations offers a powerful and scalable solution for cytoarchitectonic parcellation.

\acknowledgments 
This work was supported by the STI 2030-Major Project(2022ZD0205203 to F.X.) and “Brain Science and Brain-Inspired Intelligence”, under the subproject “New Technologies for Whole-Brain-Scale Neuronal Mesoscopic Atlas” (2022ZD0211904 to P.-C.Z)

\bibliography{report} 
\bibliographystyle{spiebib} 

\end{document}